\documentclass[aps,prc,twocolumn,superscriptaddress,showpacs]{revtex4}
\usepackage{graphicx}
\usepackage{dcolumn}
\usepackage{bm}
\usepackage{subfigure}
\usepackage{amssymb}
\newcommand{\beq}{\begin{equation}}
\newcommand{\eeq}{\end{equation}}
\newcommand{\beqar}{\begin{eqnarray}}
\newcommand{\eeqar}{\end{eqnarray}}
\newcommand{\ds}{\displaystyle}
\begin{document}

\title{Jets and decays of resonances: Two mechanisms responsible for
reduction of the elliptic flow at the CERN Large Hadron Collider (LHC), 
and restoration of constituent quark scaling}


\author{G.~Eyyubova}
\altaffiliation[Also at ]{
Skobeltsyn Institute of Nuclear Physics,
Moscow State University, RU-119991 Moscow, Russia
\vspace*{1ex}}
\affiliation{
Department of Physics, University of Oslo, PB 1048 Blindern,
N-0316 Oslo, Norway
\vspace*{1ex}}
\author{L.V.~Bravina}
\affiliation{
Department of Physics, University of Oslo, PB 1048 Blindern,
N-0316 Oslo, Norway
\vspace*{1ex}}
\author{V.L.~Korotkikh}
\affiliation{
Skobeltsyn Institute of Nuclear Physics,
Moscow State University, RU-119991 Moscow, Russia
\vspace*{1ex}}
\author{ I.P.~Lokhtin}
\affiliation{
Skobeltsyn Institute of Nuclear Physics,
Moscow State University, RU-119991 Moscow, Russia
\vspace*{1ex}}
\author{L.V.~Malinina}
\altaffiliation[Also at ]{
Department of Physics, University of Oslo, Post Box 1048 Blindern,
N-0316 Oslo, Norway and
Joint Institute for Nuclear Researches, Dubna, Moscow
Region, RU-141980, Russia
\vspace*{1ex}}
\affiliation{
Skobeltsyn Institute of Nuclear Physics,
Moscow State University, RU-119991 Moscow, Russia
\vspace*{1ex}}
\author{S.V.~Petrushanko}
\affiliation{
Skobeltsyn Institute of Nuclear Physics,
Moscow State University, RU-119991 Moscow, Russia
\vspace*{1ex}}
\author{A.M.~Snigirev}
\affiliation{
Skobeltsyn Institute of Nuclear Physics,
Moscow State University, RU-119991 Moscow, Russia
\vspace*{1ex}}
\author{E.~Zabrodin}
\altaffiliation[Also at ]{
Skobeltsyn Institute of Nuclear Physics,
Moscow State University, RU-119991 Moscow, Russia
\vspace*{1ex}}
\affiliation{
Department of Physics, University of Oslo, PB 1048 Blindern,
N-0316 Oslo, Norway
\vspace*{1ex}}

\date{\today}

\begin{abstract}
The formation and evolution of the elliptic flow pattern in Pb + Pb 
collisions at $\sqrt{s}=5.5${\it A}~TeV and in Au + Au collisions 
at $\sqrt{s}=200${\it A}~GeV are analyzed for different hadron 
species within the framework of the HYDJET++ Monte Carlo model. The 
model contains both hydrodynamic state and jets, thus allowing for 
a study of the interplay between the soft and hard processes. It is 
found that jets terminate the rise of the elliptic flow with 
increasing transverse momentum. Since jets are more influential at the
Large Hadron Collider (LHC) than at the Relativistic Heavy Ion 
Collider (RHIC), the elliptic flow at LHC should be weaker than that 
at RHIC. The influence of resonance decays on particle elliptic flow 
is also investigated. These final state interactions enhance the 
low-$p_T$ part of the $v_2$ of pions and light baryons and work 
toward the fulfillment of idealized constituent quark scaling.  
\end{abstract}


\maketitle
\section{Introduction}
\label{sec1}

In noncentral collisions between two nuclei the beam direction and 
the impact parameter vector define a reaction plane for each event. 
The observed particle yield versus azimuthal angle with respect to the 
event-by-event reaction plane might carry information on the early 
collision dynamics \cite{Ollit,Sorge}. An initial nuclear overlap 
region has an ``almond'' shape at nonzero impact parameter. If the 
produced matter further interacts and quickly thermalizes, the 
pressure that is built up within the almond-shaped region, develops 
anisotropic pressure gradients. This pressure pushes against the 
outside vacuum and the matter expands collectively. The result is an 
anisotropic azimuthal angle distribution of the detected particles. 
One can expand this azimuthal angle distribution in Fourier series
\cite{Voloshin,voloshin_2}. The second coefficient of the expansion 
$v_2$ is called elliptic flow:
\beq
\ds
v_2 \equiv \langle \cos{2 \phi} \rangle = \left\langle 
\frac{p_x^2 - p_y^2}{p_x^2 + p_y^2} \right\rangle \ .
\label{eq1}
\eeq
Here, $\phi$ is the azimuthal angle of a particle relative to the
reaction plane, and $p_x$ and $p_y$ are the in-plane and out-of-plane
components of the transverse momentum of a particle, respectively.  

It was found \cite{Sorge, Kolb} that anisotropic flow is a 
self-quenching phenomenon since it reduces spatial anisotropy as it 
evolves. Therefore, the observed elliptic flow must originate at 
early stages of the collision when the anisotropy is still present 
in the system, and no flow is generated when the spherical symmetry 
is restored. Thus, the elliptic flow can provide information about 
hot and dense matter created in relativistic heavy-ion collisions.

Although the elliptic flow has been extensively studied both
theoretically and experimentally (see, e.g., Refs.~\cite{VPS_08,Sor_09} 
and references therein), its behavior in Pb + Pb collisions 
at Large Hadron Collider (LHC) energy $\sqrt{s} = 5.5${\it A}~TeV, 
compared with that at the Relativistic Heavy Ion Collider (RHIC), 
$\sqrt{s} = 200${\it A}~GeV, remains completely unclear (for a recent 
review see Ref.~\cite{Arm_09}). Some models predict further increase 
of the $v_2$ at LHC, while others favor a similarity, $v_2^{\rm LHC} 
\approx v_2^{\rm RHIC}$, or even a decrease, $v_2^{\rm LHC} < v_2^{
\rm RHIC}$. The difference among the predictions of different models 
comes from the treatment of various processes at partonic and hadronic 
levels, such as equilibration, features of quark-hadron phase 
transition, hadronic cascade, equation of state (EOS), and cross 
section of partonic scattering, as well as from the initial conditions. 
To the best of our knowledge, the interplay between the ideal 
hydrodynamic behavior and jets, that is, the influence of the jets on 
hydrodynamic flow with rising collision energy, has not been elaborated 
yet. For this purpose we employ the HYDJET++ model \cite{hydjet++} to 
estimate the azimuthal anisotropy of particles in Pb + Pb collisions at 
$\sqrt{s} = 5.5${\it A}~TeV and compare the obtained results with 
calculations for Au + Au collisions at the top RHIC energy $\sqrt{s} = 
200${\it A}~GeV and also with the experimental data.

The HYDJET++ model \cite{hydjet++} is a superposition of soft and hard 
parts. These parts are independent and their contribution to the total 
multiplicity production depends on collision energy and centrality and 
is tuned by model parameters. The hard part of the model is identical 
to that of the HYDJET model \cite{hydjet} and can account for jet 
quenching effect and the shadowing effect \cite{Tyw_07}. The soft part 
of the HYDJET++ event represents the ``thermal" hadronic state where 
multiplicities are determined under assumption of thermal equilibrium 
\cite{fastmc1,fastmc2}. Hadrons are produced on the hypersurface 
represented by a parametrization of relativistic hydrodynamics with 
given freeze-out conditions. The model is capable of simultaneously 
reproducing the main features of heavy-ion collisions at RHIC, namely, 
(i) hadron spectra and ratios, (ii) radial and elliptic flow, (iii) 
femtoscopic momentum correlations, as well as (iv) high-$p_T$ hadron 
spectra. 

As the fireball expands, its temperature and energy density drop. 
Finally, at the freeze-out stage the system breaks up into hadrons 
and their resonances. The effect of resonance decays, namely, final 
state interactions, on the resulting elliptic flow of particles is 
quite important at both LHC and RHIC energies. Here, we are going to
benefit from the rich table of baryon and meson resonance states
implemented in HYDJET++.

The paper is organized as follows. Features of the model are presented
in Sec.~\ref{sec2}. In Sec.~\ref{sec3} the influence of jets on the
elliptic flow of the most abundant hadrons, such as pions, protons, 
kaons and lambdas, is studied. We show that the superposition of 
hydrodynamic flow and jets leads to reduction of the flow at high 
transverse momenta. Compared with RHIC, the jet fraction at LHC starts 
to dominate at smaller $p_T$, thus effectively decreasing the elliptic
flow of all particles. In Section~\ref{sec4} we describe the study of 
the influence of resonance decays on the $v_2$ of stable (with respect 
to strong interaction) particles. At both RHIC and LHC, the effect is
found to be the strongest for protons and lambdas, relatively moderate 
for pions, and almost absent for kaons. While jets modify the
(semi)hard part of the $v_2 (p_T)$ spectra, the contributions from
resonance decays alter their soft part. In Sec.~\ref{sec5} we show
that these contributions can account for better realization of 
the constituent quark scaling of the hadron elliptic flow. Finally,
conclusions are drawn in Sec.~\ref{sec6}.

\section{The HYDJET++ generator}
\label{sec2}

HYDJET++ is a Monte Carlo event generator for the simulation of 
relativistic heavy-ion {\it A + A} collisions as the superposition of 
a soft hydro-type state and a hard multiparton state. Both states are 
treated independently. HYDJET++~\cite{hydjet++} is a further 
development of its predecessors: the HYDJET \cite{hydjet} and FASTMC 
\cite{fastmc1,fastmc2} Monte Carlo generators. The soft part is based 
on a hydrodynamical parametrization of the initial state providing the 
thermal hadronic state generated on the chemical (\emph{single 
freeze-out} scenario) or thermal (\emph{thermal freeze-out} scenario) 
freeze-out hypersurfaces represented by a parametrization of 
relativistic hydrodynamics with given freeze-out conditions 
\cite{fastmc1,fastmc2}. The mean multiplicity of hadron species 
crossing the spacelike freeze-out hypersurface is calculated using 
effective thermal volume approximation. Note that unlike FASTMC in 
HYDJET++ the value of effective volume of the fireball $V_{\rm eff}$  
is generated for each event separately. $V_{\rm eff}$ is proportional
to the mean number of participating nucleons at the considered 
centrality (impact parameter $b$), which is calculated from the
generalization of the Glauber multiple scattering model to the case 
of independent inelastic nucleon-nucleon collisions. In the case of 
the \emph{thermal freeze-out} scenario the system expands 
hydrodynamically with frozen chemical composition, cools down and 
finally decays at a thermal freeze-out hypersurface \cite{fastmc1}. 
The two- and three-body decays of the resonances with branching ratios 
are taken from the SHARE particle decay table~\cite{share}.

The model for the hard multiparton part of the HYDJET++ event is the 
same as that for the HYDJET event generator. A detailed description of 
the physics framework of this model can be found in Ref.~\cite{hydjet}. 
The approach for multiple scattering of hard partons in dense QCD 
matter, that is, quark-gluon plasma, is based on an accumulating energy 
loss, the gluon radiation, and collisional loss being associated with 
each parton scattering in the expanding quark-gluon fluid.

The routine for generation of a single hard $NN$ collision, 
PYQUEN \cite{hydjet,pyquen}, is constructed as a modification of the 
jet event obtained with the generator of hadron-hadron interactions, 
PYTHIA$\_$6.4~\cite{pythia}. The event-by-event simulation procedure 
in PYQUEN includes the generation of the initial parton spectra with 
PYTHIA and production vertexes at the given impact parameter, 
rescattering-by-rescattering simulation of the parton path length in 
a dense zone, radiative and collisional energy loss, and final 
hadronization with the Lund string model for hard partons and 
in-medium emitted gluons. Then, the full hard part of the event 
includes PYQUEN multijets generated around its mean value according 
to the binomial distribution. The mean number of jets produced in 
{\it A + A} events is a product of the number of binary $NN$ 
subcollisions at a given impact parameter and the integral cross 
section of the hard process in $NN$ collisions with the minimal 
transverse momentum transfer, $p_T^{\rm min}$. Further details of the 
model can be found in Refs.~\cite{hydjet++,hydjet,fastmc1,fastmc2}. 

\section{How jets can diminish the elliptic flow at LHC compared 
with RHIC}
\label{sec3}

\begin{figure}
 \resizebox{\linewidth}{!}{
\includegraphics{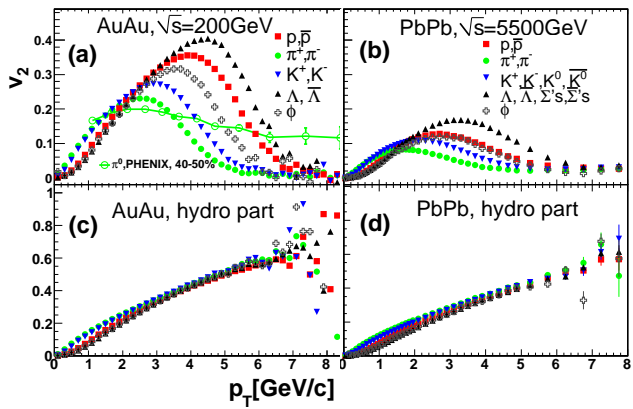}
}
\caption{(Color online) The $p_T$ dependence of total elliptic flow 
(upper row) and its hydro component (bottom row) in the HYDJET++ model
for different hadron species. Left column: Au + Au collisions at 
$\sqrt{s}=200${\it A}~GeV. Right column: Pb + Pb collisions at 
$\sqrt{s}=5.5${\it A}~TeV. Centrality for both reactions is fixed
at $c = \sigma/\sigma_{\rm geo} = 42\%$. Data for $\pi^0$ flow, shown
in (a), are taken from Ref.~\protect\cite{pi0_phenix}.
\label{v2_pt_id} }
\end{figure}

In the following we consider heavy-ion collisions at fixed centrality 
$\sigma/\sigma_{\rm geo} = \left[ b / (2 R_A)^2 \right] = 42\%$, 
corresponding to impact parameter $b\approx 1.3 R_{\rm A}$. At this 
centrality the elliptic flow is already quite strong, and the total 
hadron multiplicity is still high enough to reduce fluctuations in the 
high-$p_T$ region of particle spectra. Figure~\ref{v2_pt_id} shows the 
transverse momentum dependence of elliptic flow coefficient for the 
most abundant hadron species, namely, pions, kaons, protons, lambdas 
and sigmas, produced in 1,000,000 gold-gold collisions at $\sqrt{s} = 
200${\it A}~GeV and in ca. 500,000 lead-lead interactions at $\sqrt{s} 
= 5.5${\it A}~TeV. At least three features should be mentioned here. 
First, for both reactions, the initial increase of the flow with rising 
$p_T$ is accompanied by the rapid falloff at high values of the 
transverse momentum. Second, the pronounced feature of the RHIC 
experimental data \cite{PHENIX} reproduced by HYDJET++ in 
Fig.~\ref{v2_pt_id} is the 
\begin{figure}
 \resizebox{\linewidth}{!}{
\includegraphics{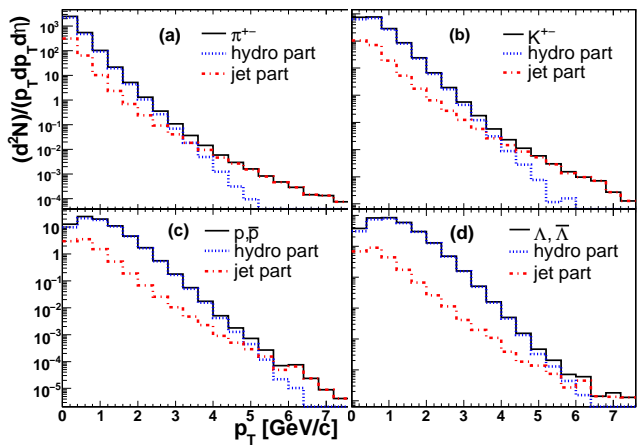}
}
\caption{(Color online) The $p_T$ distribution for different hadron 
species in HYDJET++ calculations of Pb + Pb collisions at $\sqrt{s} = 
5.5${\it A}~TeV with centrality $c = 42\%$.
\label{pt_spectr} }
\end{figure}
crossing of baryon and meson branches. Finally, the flow at LHC is 
almost 50\% lower than the flow at RHIC. We also plot recent data
concerning the elliptic flow of neutral pions at RHIC \cite{pi0_phenix}
onto the model predictions in Fig.~\ref{v2_pt_id}(a). The experimental 
data indicate a slight decrease of the $v_2$ at $p_T \geq 2$\,GeV/$c$ 
and its possible saturation at the 10\% level, whereas in the HYDJET++ 
calculations the flow in the high-$p_T$ domain is much weaker and does 
not exceed 2--3\%. Therefore, the model predictions of the flow 
excitation functions $v_2(p_T)$ above $p_T \approx 4$\,GeV/$c$ should 
be considered as qualitative ones. However, what is the origin of the 
flow drop and why is the $v_2$ at LHC significantly smaller than the 
RHIC flow? These peculiarities in the behavior of the elliptic flow 
can be explained by the interplay between the hydrolike part of the 
spectra and the jets.

The flow of the hydro part rises monotonically up to $v_2\simeq 0.5$ 
at $p_T \simeq 6$\,GeV/$c$, while the relative contribution of the soft
processes to the total particle multiplicity decreases with rising 
$p_T$, so the particles with $p_T \gtrsim 6$ GeV/$c$ are produced 
merely via hard processes, that is, jets. The jets themselves indicate 
some amount of flow due to the jet quenching effect. [The energy 
loss of the high-$p_T$ partons depends on their passing length in the 
anisotropic matter, thus giving the large yield of the high-$p_T$ 
partons in the short in-plane $(x,z)$ direction compared with that in
the long out-of plane $(y,z)$ one.] However, their flow is small and 
does not exceed 5\% in HYDJET++ calculations even at full LHC energy. 
Therefore, the superposition of the hydro part and the jets results in 
weakening of the elliptic flow after a certain transverse momentum. At 
LHC, jets turn to dominate over the soft processes at much lower values 
of $p_T$ than at RHIC, thus effectively decreasing the elliptic flow, 
$v_2^{\rm LHC} < v_2^{\rm RHIC}$ at $p_T \geq 3$\,GeV/$c$. It is worth 
mentioning that the last result stems from the assumption of 
similarity of model parameters responsible for the correct description 
of particle chemical and thermal freeze-out at both energies. (We 
have considered a naive ''scaling" of the existing physical picture of
heavy-ion interactions over the two orders of magnitude in 
center-of-mass energy to the maximum LHC energy $\sqrt{s} = $5.5\,
{\it A}~TeV. The linear extrapolation of the model parameters in 
$\log(\sqrt{s})$ to the LHC can be found in Ref.~\cite{LHC_pred}.) 
These parameters are listed in Table~\ref{tab1}. Some of 
\begin{table}
\caption{The input parameters for the HYDJET++ generator for the two 
reactions in question: 
$T_{\rm ch}$, temperature at chemical freeze-out;
$T_{\rm th}$, temperature at thermal freeze-out;
$\mu_{\rm B}$, baryon chemical potential;
$R_{\rm tran}^{\rm max}$, maximum transverse radius at thermal 
freeze-out;
$y_{\rm long}^{\rm max}$, maximum longitudinal flow rapidity at thermal 
freeze-out;
$y_{\rm tran}^{\rm max}$, maximum transverse flow rapidity at thermal
freeze-out; and
$p_T^{\rm min}$, minimum transverse momentum of parton-parton 
scattering.
For Au + Au, $\sqrt{s}=200${\it A}~GeV. 
For Pb + Pb, $\sqrt{s}=5.5${\it A}~TeV. }
\label{tab1}
\begin{center}
\begin{tabular}{lcc}
\hline \hline
 & Au + Au  &  Pb + Pb  \\
\hline
$T_{\rm ch}$                     & 165 MeV      & 170 MeV    \\ 
$T_{\rm th}$                     & 100 MeV      & 130 MeV    \\ 
$\mu_{\rm B}$                    & 28.5 MeV     & 0          \\
$R_{\rm tran}^{\rm max}$         & 10 fm        & 11 fm      \\
$y_{\rm long}^{\rm max}$         & 3.3          & 4          \\
$y_{\rm tran}^{\rm max}$         & 1.1          & 1.1        \\
$p_T^{\rm min}$                  & 3.4 GeV/$c$  & 7 GeV/$c$  \\
\hline \hline
\end{tabular}
\end{center}
\end{table}
them will probably be modified at LHC more seriously compared with the 
estimated values. Nevertheless, even if the elliptic flow at LHC in the 
low-$p_T$ region will be stronger than that at RHIC, our condition 
$v_2^{\rm LHC} < v_2^{\rm RHIC}$ will be accomplished at slightly 
higher transverse momenta, say, $p_T \geq 4$\,GeV/$c$.  

Last but not least, in HYDJET++ at low $p_T$ the elliptic flow is 
strictly ordered by particle masses. Light particles, such as pions 
and kaons, have larger flow than heavier ones, such as protons 
and lambdas. On the other hand, the slope of the $p_T$ spectra of 
light particles is steeper than that of the heavy particles, as can
be clearly seen in Fig.~\ref{pt_spectr}. As a result, the hydro 
component of the transverse momentum distribution of heavy hadrons 
dominates until larger values of $p_T$. For instance, for pions 
and kaons, it determines the spectrum up to $p_T \sim 4$\,GeV/$c$, 
whereas for protons the hydro part dominates until $p_T \sim 
5$\,GeV/$c$. Because of this, the fall of the elliptic flow with rising 
transverse momentum occurs for light particles at smaller $p_T$, and, 
at $p_T \geq 4$\,GeV/$c$, the mass ordering of the $v_2(p_T)$ spectra 
is reversed: heaviest particles possess the largest flow.

\section{The influence of resonance decays} 
\label{sec4}

At RHIC energies the transition to meson-rich matter has been found. 
As was predicted by Hagedorn \cite{Hag_65}, at high energies most of 
the particles will be produced through resonance decays with shifting 
of the average mass to a heavier sector. Table~\ref{tab2} shows the 
\begin{table}
\caption{Yields of the particles produced directly and via the 
resonance decays at the midrapidity region. Feed-down from weak decays 
of strange particles is included. For Pb + Pb, $c = 42\%$ and
$\sqrt{s}=5.5${\it A}~TeV. For Au + Au, $c = 42\%$ and 
$\sqrt{s}=200${\it A}~GeV.}
\label{tab2}
\begin{center}
\begin{tabular}{lccccc}
\hline\hline
    & $\pi^{\pm}$ & $ K(\bar K) $& $p(\bar p)$ & $\Lambda(\bar 
\Lambda) + \Sigma(\bar \Sigma)$ & $ \phi $         \\
  \hline
Pb + Pb       &       &       &      &      &      \\
all           & 860   & 185   & 63.8 & 42.3 & 6.55 \\
direct        & 169   & 81.4  & 18.6 & 14.2 & 6.5  \\
direct(\%)    & 20    & 44    & 30   & 39   & 99   \\
Au + Au       &       &       &      &      &      \\
all           & 190   & 21.5  & 13.8 & 6.6  & 1.44 \\ 
direct        & 42.9  & 10.4  & 3.2  & 1.2  & 1.43 \\
direct(\%)    & 22.5  & 48    & 22   & 18   & 99   \\ 
\hline\hline
\end{tabular}
\end{center}
\end{table}
contributions of direct and resonant production for various hadron 
species including feed-down from weak decays for the Pb + Pb event 
sample generated with HYDJET++ at the top LHC energy. One can see 
that about 80\% of pions, 70\% of protons, 60\%  of $\Sigma$ and 
$\Lambda$ hyperons and more than 50\% of kaons are produced from 
the decays of resonances.

\begin{figure}
 \resizebox{\linewidth}{!}{
\includegraphics{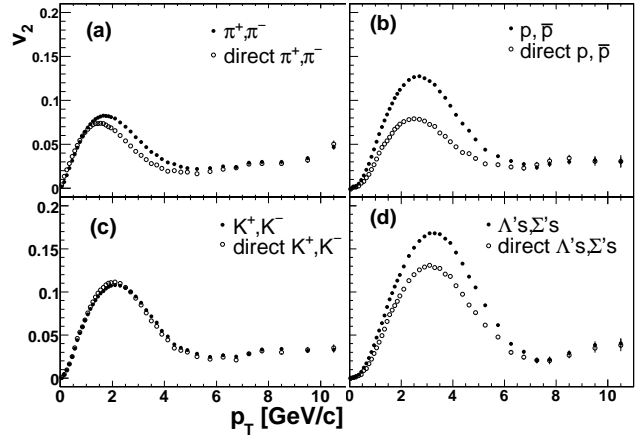}
}
\caption{The $p_T$ dependence of elliptic flow of direct hadrons 
(open symbols) and of all hadrons (full symbols) produced in the
HYDJET++ model for Pb + Pb collisions at $\sqrt{s}=5.5${\it A}~TeV 
with centrality $c = 42\%$: (a) protons, (b) pions, (c) kaons, and 
(d) lambdas plus sigmas.
\label{v2_resonance_id_lhc} }
\end{figure}
\begin{figure}
 \resizebox{\linewidth}{!}{
\includegraphics{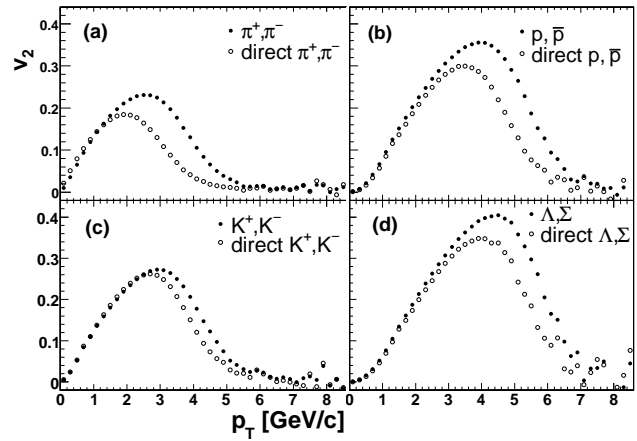}
}
\caption{The same as Fig.\protect\ref{v2_resonance_id_lhc} but for
Au + Au collisions at $\sqrt{s}=200${\it A}~GeV.
\label{v2_resonance_id_rhic} }
\end{figure}
The effect of resonance decays should be accounted for when one 
considers the development of the elliptic flow as well. The difference 
between $v_2$ of all these hadrons and $v_2$ of only directly produced 
hadrons is displayed in Figs.~\ref{v2_resonance_id_lhc} and 
\ref{v2_resonance_id_rhic} for Pb + Pb collisions at LHC and for 
Au + Au collisions at RHIC, respectively. The degree of influence of 
resonance decays on the strength of elliptic flow is quite different 
for various hadrons. The effect is strong for protons and $\Sigma + 
\Lambda$ hyperons, rather moderate for pions, and extremely small for 
kaons. For all particles, except pions, the resonances either do not 
alter the flow (kaons) or increase it for both energies in question. 
Note that in the $p_T$ region below 1\,GeV/$c$ the elliptic flow of 
direct pions is larger than the flow of all pions. This finding is in
line with the result obtained in relativistic (2+1) hydrodynamics
\cite{KH_08}. One can conclude that the soft pions emitted from the
decays of resonances have {\it lower\/} momentum anisotropy, whereas
heavier particles and hard pions demonstrate {\it larger\/} elliptic
flow of the resonances.   

\begin{figure}
\resizebox{\linewidth}{!}{
\includegraphics{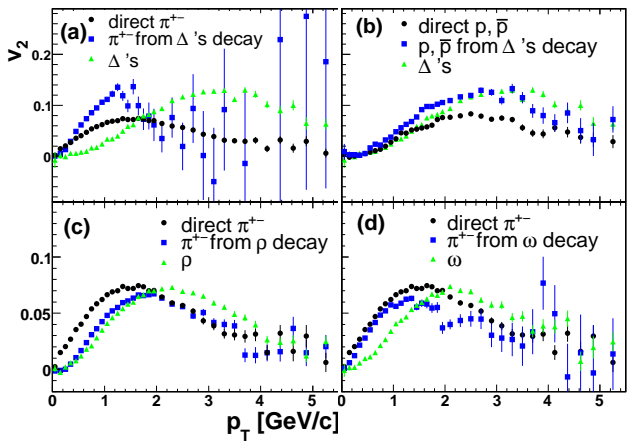}
}
\caption{(Color online) The $p_T$ dependence of elliptic flow for (a) 
charged pions and (b) protons plus antiprotons produced both directly 
(circles) and in $\Delta$ decays (squares) in the HYDJET++ model for 
Pb + Pb collisions at $\sqrt{s}=5.5${\it A}~TeV with centrality $c = 
42\%$. (c) and (d) The same as (a) but for charged pions produced in decays of 
$\rho$ and $\omega$, respectively. The flow of resonances is shown by 
triangles.
\label{v2_resonance_compare_lhc} }
\end{figure}
\begin{figure}
\resizebox{\linewidth}{!}{
\includegraphics{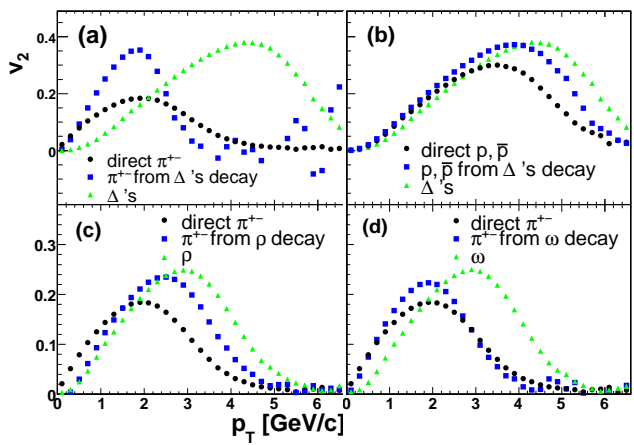}
}
\caption{(Color online) 
The same as Fig.\protect\ref{v2_resonance_compare_lhc} but for
Au + Au collisions at $\sqrt{s}=200${\it A}~GeV.
\label{v2_resonance_compare_rhic} }
\end{figure}
In order to study this peculiarity in detail, let us consider the case 
of pion and proton flow. Figures~\ref{v2_resonance_compare_lhc} and 
\ref{v2_resonance_compare_rhic} show differences in spectra of 
secondary pions and protons coming from (anti)delta ($\Delta^{++}, 
\Delta^{+}, \Delta^{0}, \Delta^{-}$) decay. When a heavy baryon 
resonance decays into a secondary baryon plus a pion, the majority of 
its transverse momentum, because of the decay kinematics, is carried 
by the baryon, while the pion is produced with lower $p_T$. 
As seen in Figs.~\ref{v2_resonance_compare_lhc}(a),
\ref{v2_resonance_compare_lhc}(b), \ref{v2_resonance_compare_rhic}(a),
and \ref{v2_resonance_compare_rhic}(b) the resulting elliptic 
flows of baryon and pion decay fractions possess similar amplitudes 
(but not the same $p_T$) as was carried by the resonance before the 
decay, that is, $\langle v_2^{\rm resonance} \rangle \approx \langle 
v_2^{\rm baryon} \rangle \approx \langle v_2^{\rm meson} \rangle$, but 
$\langle p_T^{\rm resonance} \rangle \approx \langle p_T^{\rm baryon} 
\rangle > \langle p_T^{\rm meson} \rangle$.
Since the transverse momentum spectrum of the produced pions is 
softer, the pion elliptic flow gets an extra boost at low $p_T$ from 
the intermediate-$p_T$ flow of heavy resonances see
\[ Figs.~\ref{v2_resonance_compare_lhc}(a) and 
\ref{v2_resonance_compare_rhic}(a) \]. In contrast, the secondary 
baryon has practically the same $v_2(p_T)$ dependence as that of the 
mother particle, as shown in Figs.~\ref{v2_resonance_compare_lhc}(b) 
and \ref{v2_resonance_compare_rhic}(b).

\begin{table}
\caption{Relative yields of pions and protons produced directly and via 
the resonance decays in HYDJET++. Calculations are done for Au + Au
collisions at RHIC and Pb + Pb collisions at LHC with centrality
c=42\%. For Pb + Pb, $\sqrt{s}=5.5${\it A}~TeV. For Au + Au,
$\sqrt{s}=200${\it A}~GeV.
}
\label{tab3}
\begin{center}
\begin{tabular}{ccccccc}
\hline\hline
 Hadron  & Direct & $\rho$ decay & $ K^{0}$ decay & $\omega$ decay & 
$\Lambda$ decay & $ \Delta$ decay\\
 \hline
Pb + Pb      &        &      &        &       &        &        \\
$\pi^{\pm}$  & 22\%   & 26\% &  16\%  & 11\%  &  2.3\% & 1.8\%  \\
$p,\bar p$   &30\%    &  -   &   -    &  -    & 27\%   & 15\%   \\
Au + Au      &        &      &        &       &        &        \\
$\pi^{\pm}$  & 22.5\% & 21\% & 17.8\% & 8.4\% &  2.3\% & 1.8\%  \\
$p,\bar p$   &23\%    &  -   &   -    &  -    & 30.4\% & 16\%   \\
\hline\hline
\end{tabular}
\end{center}
\end{table}
The relative contributions of different resonance channels to the
yields of protons and pions at RHIC and LHC energies are presented in
Table~\ref{tab3}. We see that, for example, lambdas and deltas alone,
together with their antistates, produce about 45\% of all protons and
antiprotons. On the other hand, the contribution of baryon resonances 
to meson spectra is relatively weak. As follows from Table~\ref{tab3}, 
many secondary pions are produced from $\rho$ and $\omega$ mesons. 
The momentum distributions of these pions are quite different, as can 
be seen in Figs.~\ref{v2_resonance_compare_lhc}(c),
\ref{v2_resonance_compare_lhc}(d), \ref{v2_resonance_compare_rhic}(c),
and \ref{v2_resonance_compare_rhic}(d).

Elliptic flow of pions from the $\rho \to \pi\pi$ decay almost 
coincides with $v_2^{\rho}$, while in the three-particle decay $\omega 
\to \pi\pi\pi$ pions are getting an obviously softer $p_T$ 
distribution, and thus their elliptic flow is transferred to the 
softer $p_T$ region compared with $v_2^{\omega}(p_T)$. Although the 
contributions to the elliptic flow of pions coming from baryon and 
meson resonances at $p_T \leq 1.5$\,GeV/$c$ effectively compensate 
each other, the resulting elliptic flow appears to be a bit lower than 
that of directly produced pions. For higher transverse momenta the 
contribution from the decays of $K$ mesons and heavy resonances 
determines the observed excess of the resulting pion flow over the 
direct pion flow.

In general, the resonance contributions sometimes increase and 
sometimes decrease the initial elliptic flow assigned to directly 
produced particles. Some hadrons, such as the $\phi$ meson, do not 
get a feed-down from the resonances, and their flow profiles remain
unchanged. Modification of the elliptic flow of pions and light 
baryons, especially pronounced at LHC energy, can lead to violation 
of the hydro-induced mass hierarchy in the $v_2(p_T)$ sector.

\section{Number-of-constituent-quark scaling}
\label{sec5}

One of the most interesting features observed in the development of 
hadron elliptic flow at RHIC is the so-called
number-of-constituent-quark (NCQ) scaling \cite{ncq_star,ncq_phen}. 
It appears that elliptic flow of any hadron species depends on the 
transverse kinetic energy $K E_T \equiv m_T - m_0$ in a similar manner 
provided both $v_2$ and $K E_T$ are divided by the number of 
constituent quarks, $n_q$, in a given hadron; thas is, $n_q = 2$ for a 
meson and $n_q = 3$ for a baryon. The observance of the NCQ scaling in 
a broad kinematic range implies the formation of elliptic flow already 
on a partonic level. Recent experimental studies based on higher 
statistics indicate that the scaling holds up only until $K E_T / n_q 
\approx 1$\,GeV \cite{PHENIX}.
\begin{figure}
\begin{center}
\resizebox{\linewidth}{!}{
\includegraphics[width=0.7\textwidth]{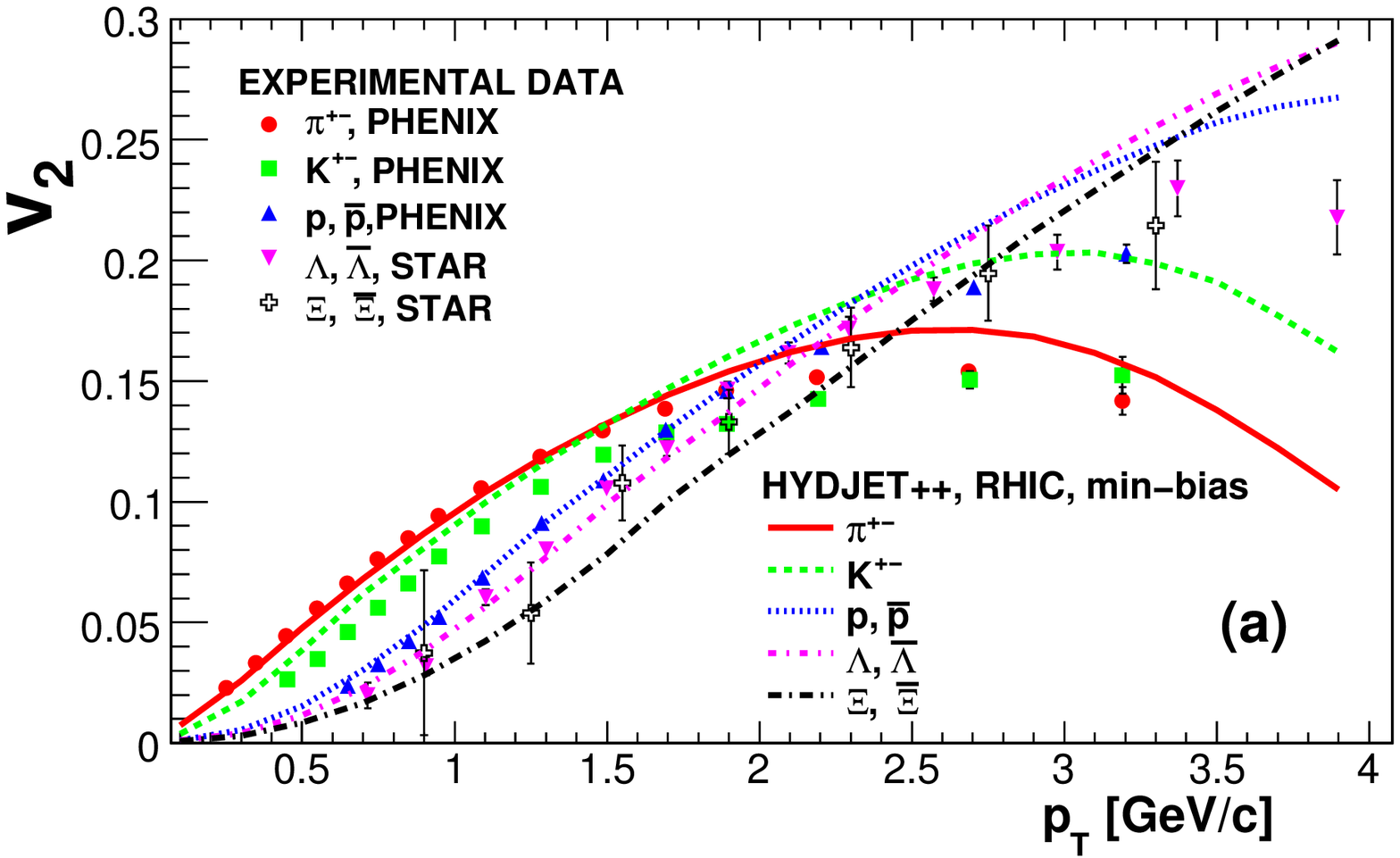}
}
\resizebox{\linewidth}{!}{
\includegraphics[width=0.7\textwidth]{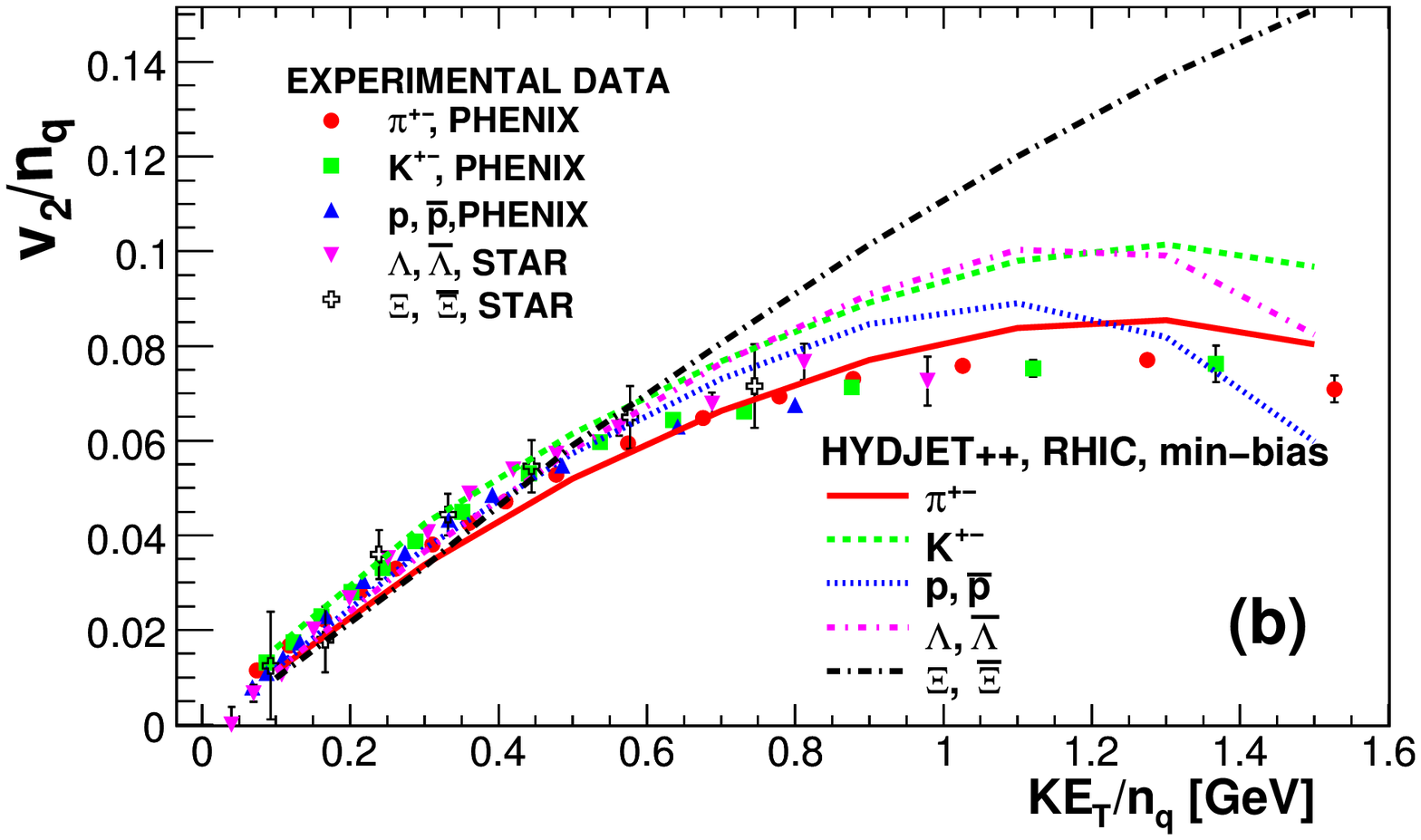}
}
\end{center}
\caption{(Color online) (a) The $p_T$ dependence of elliptic flow in 
the HYDJET++ model for different hadron species (lines) and comparison 
with RHIC data (symbols). (b) The same as (a) but for scaling 
variables, $v_2/n_q$ vs $K E_T$.
\label{rhic_data} }
\end{figure}

The experimental data on particle elliptic flow and NCQ scaling in 
1,000,000 minimum bias Au + Au collisions at $\sqrt{s} = 
200${\it A}~GeV are shown in Fig.~\ref{rhic_data} together with the 
results of the HYDJET++ simulations. Since no information about quark 
content is used in the hydro part of the HYDJET++ model, the scaling 
might not necessarily have been observed. For instance, a $\phi$ meson 
is heavier than a proton and, according to the mass ordering, its 
normalized flow $v_2^\phi (p_T)/2$ is larger than the proton flow 
$v_2^p (p_T)/3$ already at $p_T \geq 0.8$\,GeV/$c$. On the other hand, 
the elliptic flow in the model is fit to describe experimental data in 
the low-$p_T$ range, as displayed in Fig.~\ref{rhic_data}(a). 
Therefore, the dependence $v_2/n_q(KE_T/n_q)$ may also exhibit the
scaling trend. The model calculations presented in 
Fig.~\ref{rhic_data}(b) seem to obey the approximate NCQ scaling up 
to $KE_T/n_q \approx 0.7$\,GeV or even to higher values if we exclude 
heavy hyperons, such as $\Xi$ and $\bar \Xi$, from our consideration. 
(There are several experimental and theoretical lines of evidence that 
multistrange hyperons are frozen earlier; see Ref.~\cite{Xi_frees}.) 
Is this just a coincidence, and what is the role of resonance decays?

\begin{figure}
\resizebox{\linewidth}{!}{
\includegraphics{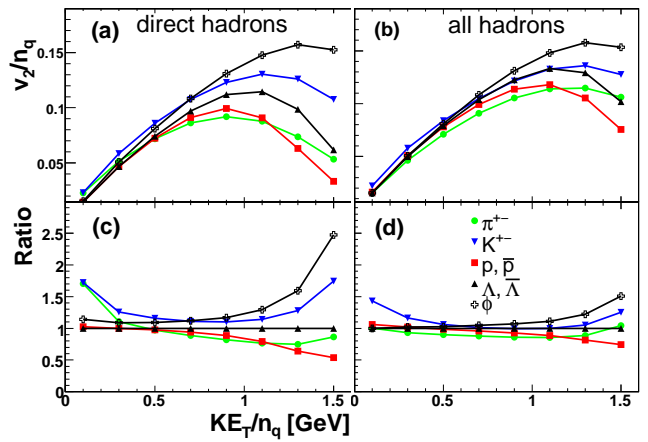}
}
\caption{(Color online) Upper row: The $KE_T/n_q$ dependence of 
elliptic flow for (a) direct hadrons and (b) all hadrons produced in 
the HYDJET++ model for Au + Au collisions at $\sqrt{s} = 
200${\it A}~GeV with centrality $c = 42\%$. Bottom row: The $KE_T/n_q$
dependence of the ratios $(v_2/n_q)\left/(v_2^\Lambda/3) \right.$ for 
(c) direct hadrons and (d) all hadrons.
\label{KEt_scaling_rhic} }
\end{figure}
To answer these questions we plot HYDJET++ excitation functions 
$v_2/n_q (KE_T/n_q)$ for direct hadrons only and for all hadrons 
produced in gold-gold collisions at top RHIC energy in 
Figs.~\ref{KEt_scaling_rhic}(a) and \ref{KEt_scaling_rhic}(b),
respectively. In order to see possible deviations from the scaling
behavior more distinctly, the ratios $v_2^{(i)}/n_q : v_2^\Lambda/3,
\ i = \pi^\pm, K^\pm, p/{\bar p}, \Lambda /{\bar \Lambda}, \phi$ are
presented in Figs.~\ref{KEt_scaling_rhic}(c) and 
\ref{KEt_scaling_rhic}(d). One may note that, for direct hadrons, the 
NCQ scaling is fulfilled within the 20\% accuracy limit in the interval 
0.2 $\leq K E_T \leq$ 0.8\,GeV. At higher and lower transverse energies 
the curves diverge, as shown in Figs.~\ref{KEt_scaling_rhic}(a) and 
\ref{KEt_scaling_rhic}(c). The situation is markedly improved after the 
resonance decays are taken into account, as demonstrated in 
Figs.~\ref{KEt_scaling_rhic}(b) and \ref{KEt_scaling_rhic}(d). All 
ratios, except for kaons at $K E_T \leq 0.2$\,GeV, are much closer to 
unity, and the scaling appears to hold at 10\% up to $K E_T \geq 
1.0$\,GeV. The spectrum and, therefore, the elliptic flow of $\phi$ 
mesons remain unchanged because of the absence of resonance feed-down. 
In contrast, protons and lambdas are significantly boosted by heavy 
resonances, and their flows are enhanced at $K E_T \geq 0.5$\,GeV.

\begin{figure}
\resizebox{\linewidth}{!}{
\includegraphics{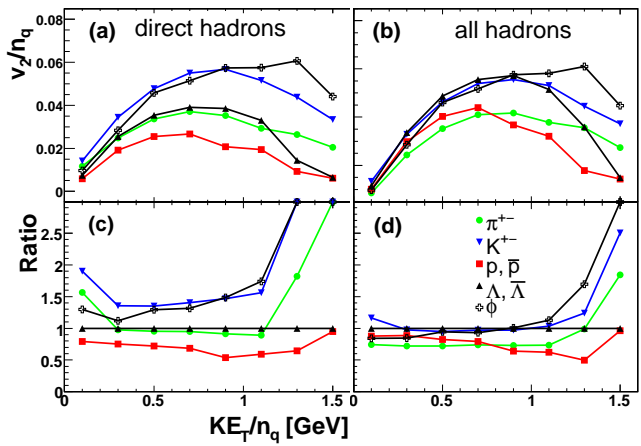}
}
\caption{(Color online) The same as Fig.\protect\ref{KEt_scaling_rhic} 
but for Pb + Pb collisions at $\sqrt{s}=5.5${\it A}~TeV.
\label{KEt_scaling_lhc} }
\end{figure}

The same mechanism also works at LHC energies. However, because of the 
influence of jets at relatively low transverse momenta, there is no 
resemblance of the NCQ scaling for directly produced hadrons, as shown
in Figs.~\ref{KEt_scaling_lhc}(a) and \ref{KEt_scaling_lhc}(c). Here, 
all curves seem to diverge in the entire transverse energy range. 
Again, the decays of resonances essentially raise up the elliptic flow 
of protons and lambdas displayed in Figs.~\ref{KEt_scaling_lhc}(b) and
\ref{KEt_scaling_lhc}(d). However, the effect of jets at LHC is 
too strong and, although all ratios of particle flows presented in 
Fig.~\ref{KEt_scaling_lhc}(d) are almost parallel within the energy
range 0.1\,$\leq KE_T \leq$\,1\,GeV, the realization of theapproximate 
number-of-constituent-quark scaling becomes worse compared with the 
RHIC case. 

A few remarks on results obtained in other calculations are in order
here. The effect of resonance decays on the elliptic flow of stable 
hadrons in Au + Au collisions at RHIC was studied in Ref.~\cite{Greco} 
in the framework of the coalescence model and in Ref.~\cite{DESX_04} 
within the parametrized thermal model. In contrast to our study, the 
authors of both works postulate that the elliptic flows of directly 
produced stable particles and resonances obey the 
number-of-constituent-quark scaling. After that, the resonances 
account for a substantial part of deviation of pion elliptic flow 
from the NCQ scaling. We do not assume the automatic performance of the 
NCQ scaling, but in our model calculations the resonance decays also 
modify the pion elliptic flow by enhancing it in the intermediate-$p_T$ 
region. In accordance with Ref.~\cite{Greco} kaon elliptic flow in our 
calculations is not affected by the decays of resonances, but we see 
the strong contribution from heavier resonances to the momentum 
anisotropies of protons and lambdas. This outlines the importance of a
complete table of resonances in any microscopic or macroscopic model 
designed for the description of heavy-ion collisions. For instance, the
present version of the HYDJET++ model employs ca. 350 baryon and meson 
states as well as their antistates.

It was suggested recently \cite{ncq_break} that NCQ scaling could be
used as a unique probe of strongly interacting partonic matter. The 
scaling holds if the hadrons are produced predominantly from the 
coalescence of quarks, whereas it is broken if the hadronization 
proceeds via the string fragmentation. We see, however, that the jet 
effects become increasingly important at ultrarelativistic energies. 
Particularly, the interplay between jets and hydrolike flow can cause 
breaking of the NCQ scaling for the particle elliptic flow at LHC. 
 
\section{Conclusions}
\label{sec6}

The elliptic flow pattern in Pb + Pb collisions at $\sqrt{s} = 
5.5${\it A}~TeV and in Au + Au collisions at $\sqrt{s} = 
200${\it A}~GeV is analyzed for different hadron species in the 
framework of the HYDJET++ Monte Carlo model. The model contains both 
parametrized hydrodynamics and jets. This allows one to study the 
interplay between hard and soft processes and to reveal their role in 
the formation and evolution of the elliptic flow. It is worth 
mentioning that the model parameters for the soft hydro component are 
fixed at RHIC energy to simultaneously describe hadron yields, energy 
spectra, anisotropic flow, and femtoscopic momentum correlations. 
These parameters are not expected to change significantly at the LHC 
energy.
 
Three general tendencies have been observed. First, at a certain 
transverse momentum, particles produced via jets start to dominate 
over the hydrolike particles. Since the flow of jet particles 
determined merely by the jet quenching is weak, the resulting flow 
$v_2(p_T)$ experiences falloff in the intermediate $p_T$ range. 
Abundant production of jets at LHC will effectively decrease the 
hadron elliptic flow at smaller values of $p_T$ compared with RHIC. 
Therefore, we expect that the elliptic flow of stable hadrons at LHC 
will be smaller than its RHIC counterpart, $v_2^{\rm LHC}(p_T) < 
v_2^{\rm RHIC}(p_T)$, at $p_T \gtrsim 3$\,GeV/$c$. 

Second, jets also account for changing of the mass ordering of the 
hadron elliptic flow at intermediate and high transverse momenta.
After the falloff the mass ordering of the $v_2(p_T)$ spectra becomes 
opposite to the initial one; that is, the heavier particle has the 
larger flow.

Third, resonances are shown to significantly modify the elliptic flows
of directly produced hadrons. The $v_2(p_T)$ of kaons is not affected
by the decays, while the flows of pions and especially protons and
lambdas are enhanced. It is the effect of resonance decays that pushes
the particle spectra toward the fulfillment of the constituent quark
number scaling. At the LHC energy the jets should alter the hadronic
elliptic flow already at intermediate transverse momenta, thus 
completely removing resemblance of the NCQ scaling for directly
produced hadrons. After the resonance decays the elliptic flows of
different hadrons move closer to each other; however, only the
approximate $K E_T/n_q$ scaling holds.   

\begin{acknowledgments}
Fruitful discussions with L.~Sarycheva, J.~Schukraft, J.~Stachel, and
R.~Snellings are gratefully acknowledged. We would like to thank 
I.~Arsene, A.~Gribushin, and K.~Tywoniuk for effective collaboration 
at different stages of the HYDJET++ development. This work was 
supported in part by the QUOTA Program, Norwegian Research Council 
(NFR), under Contract No. 185664/V30, the Russian Foundation for Basic 
Research (Grant Nos. 08-02-91001 and 08-02-92496), Grants Nos. 
107.2008.2 and 1456.2008.2 of the President of the Russian Federation, 
and the Dynasty Foundation.
\end{acknowledgments}



\begin{thebibliography}{99}

\bibitem{Ollit} J.-Y.~Ollitrault, Phys. Rev. D {\bf 46}, 229 (1992).

\bibitem{Sorge} H.~Sorge, Phys. Rev. Lett. {\bf 82}, 2048 (1999).

\bibitem{Voloshin} S.~A.~Voloshin and Y.~Zhang, 
Z. Phys. {\bf C70}, 665 (1996).

\bibitem{voloshin_2} A.~M.~Poskanzer and S.~A.~Voloshin,
Phys. Rev. C {\bf 58}, 1671 (1998).

\bibitem{Kolb} P.~F.~Kolb and U.~W.~Heinz, 
in \emph{Quark Gluon Plasma 3}, edited by R.~Hwa and X.-N.~Wang
(World Scientific, Singapore, 2003), p.634.

\bibitem{VPS_08} S.~A.~Voloshin, A.~M.~Poskanzer, and R.~Snellings,
arXiv:0809.2949 [nucl-ex].

\bibitem{Sor_09} P.~Sorensen, arXiv:0905.0174 [nucl-ex].

\bibitem{Arm_09} N.~Armesto, arXiv:0903.1330 [hep-ph].

\bibitem{hydjet++} I.~P.~Lokhtin, L.~V.~Malinina, S.~V.~Petrushanko,
A.~M.~Snigirev, I.~Arsene, and K.~Tywoniuk, 
Comput. Phys. Commun. {\bf 180}, 779 (2009).

\bibitem{hydjet} I.~P.~Lokhtin and A.~M.~Snigirev,
Eur. Phys. J. {\bf C46}, 211 (2006);
http://cern.ch/lokhtin/hydro/hydjet.html~. 

\bibitem{Tyw_07} K.~Tywoniuk, I.~C.~Arsene, L.~Bravina, A.~B.~Kaidalov, 
and E.~Zabrodin, Phys. Lett. {\bf B657}, 170 (2007).

\bibitem{fastmc1} 
N.~S.~Amelin, R.~Lednicky, T.~A.~Pocheptsov, I.~P.~Lokhtin, 
L.~V.~Malinina, A.~M.~Snigirev, Iu.~A.~Karpenko, and Yu.~M.~Sinyukov,  
Phys. Rev. C {\bf 74}, 064901 (2006).

\bibitem{fastmc2}
N.~S.~Amelin, R.~Lednicky, I.~P.~Lokhtin, L.~V.~Malinina, 
A.~M.~Snigirev, Iu.~A.~Karpenko, Yu.~M.~Sinyukov, I.~Arsene, 
and L.~Bravina, 
Phys. Rev. C {\bf 77}, 014903 (2008).

\bibitem{share}
G.~Torrieri, S.~Steinke, W.~Broniowski, W.~Florkowski, J.~Letessier,
and J.~Rafelski,
Comput. Phys. Commun. {\bf 167}, 229 (2005).

\bibitem{pyquen} http://cern.ch/lokhtin/pyquen~.

\bibitem{pythia} T.~Sjostrand, S.~Mrenna, and P.~Skands, 
J. High Energy Phys. {\bf 0605}, 026 (2006);
http://home.thep.lu.se/\verb*|~|torbjorn/Pythia.html. 

\bibitem{PHENIX} A.~Adare {\it et al.\/} (PHENIX Collaboration),
Phys. Rev. Lett. {\bf 98}, 162301 (2007).

\bibitem{pi0_phenix} R.~Wei {\it et al.\/} (PHENIX Collaboration),
arXiv:0907.0024 [nucl-ex].

\bibitem{LHC_pred} N.~Armesto (ed.) {\it et al.},
J. Phys. G {\bf 35}, 054001 (2008).

\bibitem{Hag_65} R.~Hagedorn,
Nuovo Cim. Suppl. {\bf 3}, 147 (1965).

\bibitem{KH_08} G.~Kestin and U.~Heinz,
Eur. Phys. J {\bf C61}, 545 (2009).

\bibitem{ncq_star}  J.~Adams {\it et al.\/} (STAR Collaboration),
Phys. Rev. Lett. {\bf 92}, 052302 (2004).

\bibitem{ncq_phen} S.~S.~Adler {\it et al.\/} (PHENIX Collaboration),
Phys. Rev. Lett. {\bf 91}, 182301 (2003).

\bibitem{Xi_frees} R.~Witt, J. Phys. G {\bf 34}, S921 (2007);
P.~Chaloupka, M.~Sumbera, and L.~V.~Malinina,
Acta Phys. Polon. {\bf B40}, 1185 (2009).

\bibitem{Greco} V.~Greco and C.~M.~Ko, 
Phys. Rev. C {\bf 70}, 024901 (2004).

\bibitem{DESX_04} X.~Dong, S.~Esumi, P.~Sorensen, and N.~Xu,
Phys. Lett. {\bf B597}, 328 (2004).

\bibitem{ncq_break} J.~Tian, J.~H.~Chen, Y.~G.~Ma, X.~Z.~Cai, F.~Jin, 
G.~L.~Ma, S.~Zhang, and C.~Zhong,
Phys. Rev. C {\bf 79}, 067901 (2009).
 
\end{thebibliography}
\end{document}